\documentstyle[12pt]{article}
\def\vereq#1#2{\lower3pt\vbox{\baselineskip1.5pt \lineskip1.5pt
\ialign{$#1\hfill##\hfil$\crcr#2\crcr\sim\crcr}}}
\def\alt{\mathrel{\mathpalette\vereq<}}
\def\agt{\mathrel{\mathpalette\vereq>}}

\begin{document}

\title{Hypersensitivity to perturbation: An information-theoretical 
characterization of classical and quantum chaos}

\author{R\"udiger
  Schack\thanks{E-mail: r.schack@rhbnc.ac.uk} $^{^{\hbox{\tiny(a)}}}$ 
and Carlton M. 
  Caves\thanks{E-mail: caves@tangelo.phys.unm.edu} $^{^{\hbox{\tiny(b)}}}$   
    \vspace{3mm}\\
$^{\hbox{\tiny(a)}}${\it Department of Mathematics, 
             Royal Holloway, University of London} \\
             {\it Egham, Surrey TW20 0EX, United Kingdom}  \vspace{3mm}\\
$^{\hbox{\tiny(b)}}${\it Center for Advanced Studies} \\
  {\it Department of Physics and Astronomy, University of New Mexico} \\
  {\it Albuquerque, New Mexico \thinspace 87131--1156, USA}}

\date{\today}
\maketitle

\begin{abstract}
Hypersensitivity to perturbation is a criterion for chaos based on the question
of how much information about a perturbing environment is needed to keep the
entropy of a Hamiltonian system from increasing.  In this paper we give a brief
overview of our work on hypersensitivity to perturbation in classical and
quantum systems.
\end{abstract}

\section{Introduction}

In both classical and quantum physics isolated systems can display
unpredictable behavior, but the reasons for the unpredictability are quite
different.  In classical (Hamiltonian) mechanics unpredictability is a
consequence of chaotic dynamics, or exponential sensitivity to initial
conditions, which makes it impossible to predict the phase-space trajectory of
a system to a certain accuracy from initial data given to the same accuracy.
This unpredictability, which comes from not knowing the system's initial
conditions precisely, is measured by the Kolmogorov-Sinai (KS) entropy, which
is the rate at which initial data must be supplied in order to continue
predicting the coarse-grained phase-space trajectory \cite{Alekseev1981}.  In
quantum mechanics there is no sensitivity to initial conditions in predicting
the evolution of a state vector, because the unitary evolution of quantum
mechanics preserves the inner product between state vectors.  The absence of
sensitivity to initial conditions seems to suggest that there is no quantum
chaos.  Yet quantum mechanics has an even more fundamental kind of
unpredictability, which has nothing to do with dynamics: even if a system's
state vector is known precisely, the results of measurements are generally
unpredictable.

To compare the unpredictability of classical and quantum dynamics, we first
remove the usual sources of unpredictability from consideration and then
introduce a new source of unpredictability that is the same in both classical
and quantum dynamics.  The first step is to focus in classical physics on the
evolution of phase-space distributions, governed by the Liouville equation,
instead of on phase-space trajectories, and to focus in quantum physics on the
evolution of state vectors, governed by the Schr\"odinger equation.  The
Liouville equation preserves the overlap between distributions, so there is no
sensitivity to initial conditions in predicting the evolution of a phase-space
distribution.  By shifting attention from phase-space
trajectories to distributions, we remove lack of knowledge of initial
conditions as a source of unpredictability.  Moreover, by considering only
Schr\"odinger evolution of state vectors, i.e., evolution uninterrupted by 
measurements, we eliminate the intrinsic randomness of quantum measurements 
as a source of unpredictability.

The conclusion that there is no chaos in quantum evolution is now seen to be
too facile.  Were things so simple, one would have to conclude that there is no
chaos in classical Liouville evolution either \cite{Berry1992}.  Having taken
both classical and quantum unpredictability out of the picture, we 
introduce a new source of unpredictability to investigate chaos in the
dynamics.  We do this by adding to the system Hamiltonian, either classical or
quantum mechanical, a stochastic perturbation.  We measure the unpredictability
introduced by the perturbation in terms of the increase of system entropy.  By
gathering information about the history of the perturbation, one can make the
increase of system entropy smaller.  To characterize the resistance of the
system to predictability, we compare the information gathered about the
perturbation with the entropy reduction that this information purchases.  We
say that a system is {\it hypersensitive to perturbation\/} \cite{Caves1993b}
if the perturbation information is much larger than the associated
system-entropy reduction, and we regard hypersensitivity to perturbation as the
signature of chaos in Liouville or Schr\"odinger evolution (see
Sec.~\ref{sechyp}).

For classical systems we have shown that systems with chaotic dynamics display
an {\it exponential\/} hypersensitivity to perturbation
\cite{Schack1992a,Schack1996a}, in which the ratio of perturbation information
to entropy reduction grows exponentially in time, with the exponential rate of
growth given by the KS entropy.  Thus, for classical systems, we have
established that exponential hypersensitivity to perturbation characterizes
chaos in Liouville evolution in a way that is exactly equivalent to the
standard characterization of chaos in terms of the unpredictability of
phase-space trajectories (see Sec.~\ref{secclassical}).

For a variety of quantum systems we have used numerical simulations to
investigate hypersensitivity to perturbation
\cite{Schack1993e,Schack1994b,Schack1996b}.  The simulations suggest that
hypersensitivity to perturbation provides a characterization of chaos in
quantum dynamics: quantum systems whose classical dynamics is chaotic display a
quantum hypersensitivity to perturbation, which comes about because the
perturbation generates state vectors that are nearly randomly distributed in
the system Hilbert space, whereas quantum systems whose classical dynamics is
not chaotic do not display hypersensitivity to perturbation (see
Sec.~\ref{secquantum}).

\section{Hypersensitivity to perturbation} \label{sechyp}

Hypersensitivity to perturbation, in either classical or quantum mechanics, is
defined in terms of information and entropy. The entropy $H$ of an isolated
physical system (Gibbs entropy for a classical system, von Neumann entropy for
a quantum system) does not change under Hamiltonian time evolution. If the time
evolution of the system is perturbed through interaction with an incompletely
known environment, however, averaging over the perturbation typically leads to
an entropy increase $\Delta H_{\cal S}$.  Throughout this paper, we make the
simplifying assumption that the interaction with the environment is equivalent
to a stochastic perturbation of the Hamiltonian, a restriction we hope to be
able to remove in the future. Conditions under which this assumption is valid
are discussed in \cite{Schack1996b}. The increase of the system entropy can be
limited to an amount $\Delta H_{\rm tol}$, the {\it tolerable entropy
increase}, by obtaining, from the environment, information about the
perturbation.  We denote by $\Delta I_{\rm min}$ the minimum information about
the perturbation needed, on the average, to keep the system entropy below the
tolerable level $\Delta H_{\rm tol}$. A formal definition of the quantities
$\Delta H_{\cal S}$, $\Delta H_{\rm tol}$, and $\Delta I_{\rm min}$ can be
found in~\cite{Schack1996a} for the classical case and in~\cite{Schack1996b}
for the quantum case.

Entropy and information acquire physical content in the presence of a heat
reservoir at temperature $T$. If all energy in the form of heat is ultimately
exchanged with the heat reservoir, then each bit of entropy, i.e., each bit of
{\it missing information\/} about the system state, reduces by the amount 
$k_BT\ln2$ the energy that can be extracted from the system in the form of
useful work. The connection between {\it acquired\/} information and work is
provided by Landauer's principle {\cite{Landauer1961,Landauer1988}}, according
to which not only each bit of missing information, but also each bit of
acquired information, has a free-energy cost of $k_BT\ln2$. This
cost, the {\it Landauer erasure cost}, is paid when the acquired information
is erased. Acquired information can be quantified by algorithmic information
{\cite{Chaitin1987a,Zurek1989a,Zurek1989b,Caves1990c,Schack1995a}}.

We now define that a system is hypersensitive to perturbation if the
information $\Delta I_{\rm min}$ required to reduce the system entropy from
$\Delta H_S$ to $\Delta H_{\rm tol}$ is large compared to the entropy reduction
$\Delta H_S-\Delta H_{\rm tol}$, i.e.,
\begin{equation}
{\Delta I_{\rm min}\over\Delta H_S-\Delta H_{\rm tol}}\gg1\;.
\label{eqhyp}
\end{equation}
The information $\Delta I_{\rm min}$ purchases a reduction $\Delta H_S-\Delta
H_{\rm tol}$ in system entropy, which is equivalent to an increase in the
useful work that can be extracted from the system; hypersensitivity to
perturbation means that the Landauer erasure cost of the information is much
larger than the increase in available work.

Hypersensitivity to perturbation means that the inequality~(\ref{eqhyp}) holds
for almost all values of $\Delta H_{\rm tol}$.  The inequality~(\ref{eqhyp})
tends always to hold, however, for sufficiently small values of $\Delta H_{\rm
tol}$.  The reason is that for these small values of $\Delta H_{\rm tol}$, one
is gathering enough information from the perturbing environment to track a
particular system state whose entropy is nearly equal to the initial system
entropy.  In other words, one is essentially tracking a particular realization
of the perturbation among all possible realizations.  Thus, for small values of
$\Delta H_{\rm tol}$, the information $\Delta I_{\rm min}$ becomes a property 
of the perturbation; it is the information needed to specify a particular 
realization of the perturbation.  The important regime for assessing 
hypersensitivity to perturbation is where $\Delta H_{\rm tol}$ is fairly 
close to $\Delta H_S$, and it is in this regime that one can hope that 
$\Delta I_{\rm min}$ reveals something about the system dynamics, rather 
than properties of the perturbation.

\section{Classical chaos}   \label{secclassical}

In this section we do not aim for rigor; many statements in this section are
without formal proof. Instead, our objective here is to extract the important
ideas from the rigorous analysis given in~\cite{Schack1996a} and to use
them to develop a heuristic physical picture of why chaotic systems display
exponential hypersensitivity to perturbation.  For a simple illustration and a
system where exact solutions exist, see~\cite{Schack1992a}. This section
is an abbreviated version of the discussion section of~\cite{Schack1996a}. 

Consider a classical Hamiltonian system whose dynamics unfolds on a
$2F$-dimensional phase space, and suppose that the system is perturbed by a
stochastic Hamiltonian whose effect can be described as diffusion on phase
space.  Suppose that the system is globally chaotic with KS entropy~$K$.  For
such a system a phase-space density is stretched and folded by the chaotic
dynamics, developing exponentially fine structure as the dynamics proceeds.  A
simple picture is that the phase-space density stretches exponentially in half
the phase-space dimensions and contracts exponentially in the other half of the
dimensions.

The perturbation is characterized by a perturbation strength and by correlation
cells.  We can take the perturbation strength to be the typical distance (e.g.,
Euclidean distance with respect to some fixed set of canonical co\"ordinates)
that a phase-space point diffuses under the perturbation during an $e$-folding
time, $F/K\ln2$, in a typical contracting dimension.  The perturbation becomes
effective (in a sense defined precisely in Ref.~\cite{Schack1996a}) when the
phase-space density has roughly the same size in the contracting dimensions as
the perturbation strength.  Once the perturbation becomes effective, the
effects of the diffusive perturbation and of the further exponential
contraction roughly balance one another, leaving the {\it average\/}
phase-space density with a constant size in the contracting dimensions.

The correlation cells are phase-space cells over which the effects of the
perturbation are well correlated and between which the effects of the
perturbation are essentially uncorrelated.  We assume that all the correlation
cells have approximately the same phase-space volume.  We can get a rough idea
of the effect of the perturbation by regarding the correlation cells as
receiving independent perturbations.  Moreover, the diffusive effects of the
perturbation during an $e$-folding time $F/K\ln2$ are compressed exponentially
during the next such $e$-folding time; this means that once the perturbation
becomes effective, the main effects of the perturbation at a particular time
are due to the diffusion during the immediately preceding $e$-folding time.

Since a chaotic system cannot be shielded forever from the effects of the
perturbation, we can choose the initial time $t=0$ to be the time at which the
perturbation is just becoming effective.  We suppose that at $t=0$ the
unperturbed density is spread over $2^{-Kt_0}$ correlation cells, $t_0$ being
the time when the unperturbed density occupies a single correlation cell.  The
essence of the KS entropy is that for large times $t$ the unperturbed density
spreads over
\begin{equation}
{\cal R}(t)\sim 2^{K(t-t_0)}
\end{equation}
correlation cells, in each of which it occupies roughly the same phase-space
volume.  The exponential increase of ${\cal R}(t)$ continues until the
unperturbed density is spread over essentially all the correlation cells.  We
can regard the unperturbed density as being made up of {\it subdensities}, one
in each occupied correlation cell and all having roughly the same phase-space
volume.

After $t=0$, when the perturbation becomes effective, the {\it average\/}
density continues to spread exponentially in the expanding dimensions.  As
noted above, this spreading is not balanced by contraction in the other
dimensions, so the phase-space volume occupied by the average density grows as
$2^{Kt}$, leading to an entropy increase
\begin{equation}
\Delta H_{\cal S}\sim\log_2(2^{Kt})=Kt\;.
\end{equation}
Just as the unperturbed density can be broken up into subdensities, so the
average density can be broken up into {\it average subdensities}, one in each
occupied correlation cell.  Each average subdensity occupies a phase-space
volume that is $2^{Kt}$ times as big as the volume occupied by an unperturbed
subdensity.

The unperturbed density is embedded within the phase-space volume occupied by
the average density and itself occupies a volume that is smaller by a factor of
$2^{-Kt}$.  We can picture a {\it perturbed\/} density crudely by imagining
that in each occupied correlation cell the unperturbed subdensity is moved
rigidly to some new position within the volume occupied by the {\it average\/}
subdensity; the result is a {\it perturbed subdensity}.  A {\it perturbed
density\/} is made up of perturbed subdensities, one in each occupied
correlation cell.  All of the possible perturbed densities are produced by the
perturbation with roughly the same probability.

Suppose now that we wish to hold the entropy increase to a tolerable amount
$\Delta H_{\rm tol}$.  We must first describe what it means to specify the
phase-space density at a level of resolution set by a tolerable entropy
increase $\Delta H_{\rm tol}$.  An approximate description can be obtained in
the following way.  Take an occupied correlation cell, and divide the volume
occupied by the average subdensity in that cell into $2^{\Delta H_{\cal
S}-\Delta H_{\rm tol}}$ nonoverlapping volumes, all of the same size.
Aggregate all the perturbed subdensities that lie predominantly within a
particular one of these nonoverlapping volumes to produce a {\it coarse-grained
subdensity}.  There are $2^{\Delta H_{\cal S}-\Delta H_{\rm tol}}$
coarse-grained subdensities within each occupied correlation cell, each having
a phase-space volume that is bigger than the volume occupied by a perturbed
subdensity by a factor of
\begin{equation}
{2^{Kt}\over2^{\Delta H_{\cal S}-\Delta H_{\rm tol}}}=
2^{\Delta H_{\rm tol}}\;.
\label{cgvolume}
\end{equation}
A {\it coarse-grained density\/} is made up by choosing a coarse-grained
subdensity in each occupied correlation cell.  A coarse-grained density
occupies a phase-space volume that is bigger than the volume occupied by the
unperturbed density by the factor $2^{\Delta H_{\rm tol}}$ of
Eq.~(\ref{cgvolume}) and hence represents an entropy increase
\begin{equation}
\log_2\!\left(2^{\Delta H_{\rm tol}}\right)=\Delta H_{\rm tol}\;.
\end{equation}
Thus to specify the phase-space density at a level of resolution set by $\Delta
H_{\rm tol}$ means roughly to specify a coarse-grained density.  The further
entropy increase on averaging over the perturbation is given by
\begin{equation}
\log_2\!\left(2^{\Delta H_{\cal S}-\Delta H_{\rm tol}}\right)=
\Delta H_{\cal S}-\Delta H_{\rm tol}\;.
\label{furtherincrease}
\end{equation}

What about the information $\Delta I_{\rm min}$ required to hold the entropy
increase to $\Delta H_{\rm tol}$?  Since there are $2^{\Delta H_{\cal S}-\Delta
H_{\rm tol}}$ coarse-grained subdensities in an occupied correlation cell, each
produced with roughly the same probability by the perturbation, it takes
approximately $\Delta H_{\cal S}-\Delta H_{\rm tol}$ bits to specify a
particular coarse-grained subdensity.  To describe a coarse-grained density,
one must specify a coarse-grained subdensity in each of the ${\cal R}(t)$
occupied correlation cells.  Thus the information required to specify a
coarse-grained density---and, hence, the information required to hold the
entropy increase to $\Delta H_{\rm tol}$---is given by
\begin{equation}
\Delta I_{\rm min}\sim{\cal R}(t)(\Delta H_{\cal S}-\Delta H_{\rm tol}) \;,
\label{picsum}
\end{equation}
corresponding to there being a total of $(2^{\Delta H_{\cal S}-\Delta H_{\rm
tol}})^{{\cal R}(t)}$ coarse-grained densities.  The entropy
increase~(\ref{furtherincrease}) comes from counting the number of {\it
nonoverlapping\/} coarse-grained densities that are required to fill the volume
occupied by the average density, that number being $2^{\Delta H_{\cal S}-\Delta
H_{\rm tol}}$. In contrast, the information $\Delta I_{\rm min}$ comes from
counting the exponentially greater number of ways of forming {\it
overlapping\/} coarse-grained densities by choosing one of the $2^{\Delta
H_{\cal S}-\Delta H_{\rm tol}}$ nonoverlapping coarse-grained subdensities in
each of the ${\cal R}(t)$ correlation cells.

The picture developed in this section, summarized neatly in Eq.~(\ref{picsum}),
requires that $\Delta H_{\rm tol}$ be big enough that a coarse-grained
subdensity is much larger than a perturbed subdensity, so that we can talk
meaningfully about the perturbed subdensities that lie predominantly {\it
within\/} a coarse-grained subdensity.  If $\Delta H_{\rm tol}$ becomes too
small, Eq.~(\ref{picsum}) breaks down, and the information $\Delta I_{\rm
min}$, rather than reflecting a property of the chaotic dynamics as in
Eq.~(\ref{picsum}), becomes essentially a property of the perturbation,
reflecting a counting of the number of possible realizations of the
perturbation.

The boundary between the two kinds of behavior of $\Delta I_{\rm min}$ is set
roughly by the number $F$ of contracting phase-space dimensions.  When $\Delta
H_{\rm tol}/F\agt1$, the characteristic scale of a coarse-grained subdensity in
the contracting dimensions is a factor of
\begin{equation}
\left(2^{\Delta H_{\rm tol}}\right)^{1/F}=2^{\Delta H_{\rm tol}/F}\agt2 
\end{equation}
larger than the characteristic size of a perturbed subdensity in the
contracting dimensions.  In this regime the picture developed in this section
is at least approximately valid, because a coarse-grained subdensity can
accommodate several perturbed subdensities in each contracting dimension.  The
information $\Delta I_{\rm min}$ quantifies the effects of the perturbation on
scales as big as or bigger than the finest scale set by the system dynamics.
These effects, as quantified in $\Delta I_{\rm min}$, tell us directly about
the size of the exponentially fine structure created by the system dynamics.
Thus $\Delta I_{\rm min}$ becomes a property of the system dynamics, rather
than a property of the perturbation.

In contrast, when $\Delta H_{\rm tol}/F\alt1$, we are required to keep track of
the phase-space density on a very fine scale in the contracting dimensions, a
scale smaller than the characteristic size of a perturbed subdensity in the
contracting dimensions.  Subdensities are considered to be distinct, even
though they overlap substantially, provided that they differ by more than this
very fine scale in the contracting dimensions.  The information $\Delta I_{\rm
min}$ is the logarithm of the number of realizations of the perturbation which
differ by more than this very fine scale in at least one correlation cell.  The
information becomes a property of the perturbation because it reports on the
effects of the perturbation on scales finer than the finest scale set by the
system dynamics---i.e., scales that are, at the time of interest, irrelevant to
the system dynamics.

We are now prepared to put in final form the exponential hypersensitivity 
to perturbation of systems with a positive KS entropy:  
\begin{equation}
{\Delta I_{\rm min}\over\Delta H_{\cal S}-\Delta H_{\rm tol}}\sim
{\cal R}(t)\sim 2^{K(t-t_0)}\;\;\;\mbox{for $\Delta H_{\rm tol}\agt F$.}
\label{picsum2}
\end{equation}
Once the chaotic dynamics renders the perturbation effective, this exponential
hypersensitivity to perturbation is essentially independent of the form and
strength of the perturbation.  Its essence is that within each correlation cell
there is a roughly even trade-off between entropy reduction and information,
but for the entire phase-space density the trade-off is exponentially
unfavorable because the density occupies an exponentially increasing number of
correlation cells, in each of which it is perturbed independently.

What about systems with regular, or integrable dynamics?  Though we expect no
universal behavior for regular systems, we can get an idea of the possibilities
 from the heuristic description developed in this section.  Hypersensitivity to
perturbation requires, first, that the phase-space density develop structure on
the scale of the strength of the perturbation, so that the perturbation becomes
effective, and, second, that after the perturbation becomes effective, the
phase-space density spread over many correlation cells.

For many regular systems there will be no hypersensitivity simply because the
phase-space density does not develop fine enough structure.  Regular dynamics
can give rise to nonlinear shearing, however, in which case the density can
develop structure on the scale of the strength of the perturbation and can
spread over many correlation cells.  In this situation, one expects the picture
developed in this section to apply at least approximately: to hold the entropy
increase to $\Delta H_{\rm tol}$ requires giving $\Delta H_{\cal S}-\Delta
H_{\rm tol}$ bits per occupied correlation cell; $\Delta I_{\rm min}$ is
related to $\Delta H_{\rm tol}$ by Eq.~(\ref{picsum}), with ${\cal R}(t)$ being
the number of correlation cells occupied at time $t$.  Thus regular systems can
display hypersensitivity to perturbation if ${\cal R}(t)$ becomes large
(although this behavior could be eliminated by choosing correlation cells that
are aligned with the nonlinear shearing produced by the system dynamics), but
they cannot display {\it exponential\/} hypersensitivity to perturbation
because the growth of ${\cal R}(t)$ is slower than exponential.

A more direct way of stating this conclusion is to reiterate what we have
explained in this section and shown in Ref.~\cite{Schack1996a}: Exponential
hypersensitivity to perturbation is equivalent to the spreading of phase-space
densities over an exponentially increasing number of phase-space cells; such
exponential spreading holds for chaotic, but not for regular systems and is
quantified by a positive value of the Kolmogorov-Sinai entropy.

\section{Quantum chaos}   \label{secquantum}

\subsection{Distribution of vectors in Hilbert space} \label{secdist}

The simplifying restriction on the interaction with the environment made in
Sec.~\ref{sechyp} means, for the quantum case, that the interaction with the
environment is equivalent to a stochastic unitary time evolution. Given this
assumption, we can proceed as follows.  At a given time, we describe the result
of the perturbed time evolution by a list ${\cal L}=
(|\psi_1\rangle,\ldots,|\psi_N\rangle)$ of $N$ vectors in $D$-dimensional
Hilbert space, with probabilities $q_1,\ldots,q_N$, each vector in the list
corresponding to a particular realization of the perturbation, which we call a
{\it perturbation history}.  Averaging over the perturbation leads to a
system density operator
\begin{equation}
\hat\rho_{\cal S}=\sum_{j=1}^N q_j |\psi_j\rangle\langle\psi_j|\;,
\label{eqrhos}
\end{equation}
with entropy 
\begin{equation}
\Delta H_{\cal S} = 
-{\rm tr} \Bigl(\hat\rho_{\cal S} \log_2\hat\rho_{\cal S} \Bigr) \;.
\label{eqdelhs}
\end{equation}
Consider 
the class of measurements on the environment whose outcomes partition the 
list $\cal L$ into $R$ groups labeled by $r=1,\ldots,R$.  We denote by 
$N_r$ the number of vectors in the $r$th group ($\sum_{r=1}^{R}N_r=N$). 
The $N_r$ vectors in the $r$th group and their probabilities are denoted by
$|\psi^r_1\rangle,\ldots,|\psi^r_{N_r}\rangle$ and $q^r_1,\ldots,q^r_{N_r}$,
respectively. The measurement outcome $r$, occurring with probability
\begin{equation}
p_r=\sum_{i=1}^{N_r}q^r_i \;,
\label{eqpr}
\end{equation}
indicates that the system state is in the $r$th group. The system 
state conditional on the measurement outcome $r$ is described by the 
density operator
\begin{equation}
\hat\rho_r=p_r^{-1}\sum_{i=1}^{N_r}q_i^r|\psi^r_i\rangle\langle\psi^r_i|\;.
\label{eqrhor}
\end{equation}
We define the conditional system entropy
\begin{equation}
\Delta H_r = -{\rm tr} \Bigl( \hat\rho_r \log_2 \hat\rho_r \Bigr) \;,
\label{eqdelhr}
\end{equation}
the average conditional entropy
\begin{equation}
\Delta H = \sum_r p_r \Delta H_r \;,
\end{equation}
and the average information
\begin{equation}
\Delta I = -\sum_r p_r \log_2 p_r \;.
\label{eqdeli}
\end{equation}

We now describe nearly optimal measurements, i.e., nearly optimal groupings,
for which $\Delta I$ is a close approximation to $\Delta I_{\rm min}$, the
minimum information about the environment needed, on the average, to keep the
system entropy below a given tolerable entropy $\Delta H_{\rm tol}$, as
described in Sec.~\ref{sechyp}.  Given $\Delta H_{\rm tol}$, we want to
partition the list of vectors $\cal L$ into groups so as to minimize the
information $\Delta I$ without violating the condition $\Delta H\leq\Delta
H_{\rm tol}$.  To minimize $\Delta I$, it is clearly favorable to make the
groups as large as possible. Furthermore, to reduce the contribution to $\Delta
H$ of a group containing a given number of vectors, it is favorable to choose
vectors that are as close together as possible in Hilbert space.  Here the
distance between two vectors $|\psi_1\rangle$ and $|\psi_2\rangle$ can be
quantified in terms of the Hilbert-space angle \cite{Wootters1981}
\begin{equation}
\phi = \cos^{-1}\Bigl(|\langle\psi_1|\psi_2\rangle|\Bigr) \;.
\end{equation}
Consequently, to find a nearly optimal grouping, we choose an arbitrary {\it
resolution angle\/} $\phi$ ($0\leq\phi\leq\pi/2$) and group together vectors
that are less than an angle $\phi$ apart.  More precisely, groups are formed in
the following way.  Starting with the first vector, $|\psi_1\rangle$, in the 
list $\cal L$, the first group is formed of $|\psi_1\rangle$ and all vectors
in $\cal L$ that are within an angle $\phi$ of $|\psi_1\rangle$. The same
procedure is repeated with the remaining vectors to form the second group, then
the third group, continuing until no ungrouped vectors are left.  This grouping
of vectors corresponds to a partial averaging over the perturbations. To
describe a vector at resolution level $\phi$ amounts to averaging over those
details of the perturbation that do not change the final vector by more than an
angle $\phi$.

For each resolution angle $\phi$, the grouping procedure described above
defines an average conditional entropy $\Delta H\equiv\Delta H(\phi)$ and an
average information $\Delta I\equiv\Delta I(\phi)$. If we choose, for a given
$\phi$, the tolerable entropy $\Delta H_{\rm tol}=\Delta H(\phi)$, then to a
good approximation, the information $\Delta I_{\rm min}$ is given by $\Delta
I_{\rm min}\simeq\Delta I(\phi)$.  By determining the entropy $\Delta H(\phi)$
and the information $\Delta I(\phi)$ as functions of the resolution angle
$\phi$, there emerges a rather detailed picture of how the vectors are
distributed in Hilbert space.  If $\Delta I(\phi)$ is plotted as a function of
$\Delta H(\phi)$ by eliminating the angle $\phi$, one obtains a good
approximation to the functional relationship between $\Delta I_{\rm min}$ and
$\Delta H_{\rm tol}$.

As a further characterization of our list of vectors, we calculate the
distribution $g(\phi)$ of Hilbert-space angles
$\phi=\cos^{-1}(|\langle\psi|\psi'\rangle|)$ between all pairs of vectors
$|\psi\rangle$ and $|\psi'\rangle$. For vectors distributed randomly in
$D$-dimensional Hilbert space, the distribution function $g(\phi)$ is given by
{\cite{Schack1994b}}
\begin{equation}
g(\phi)=2(D-1)(\sin\phi)^{2D-3}\cos\phi \;.
\end{equation}
The maximum of this $g(\phi)$ is located at 
$\phi=\arccos\!\left(\sqrt{2(D-1)}\,\right)$; for
large-dimensional Hilbert spaces, $g(\phi)$ is very strongly peaked near the
maximum, which is located at $\phi\simeq\pi/2-1/\sqrt{2D}$, very near
$\pi/2$.

To investigate if a quantum map shows hypersensitivity to perturbation, we use
the following numerical method. We first compute a list of vectors
corresponding to different perturbation histories. Then, for about 50 values of
the angle $\phi$ ranging from 0 to $\pi/2$, we group the vectors in the nearly
optimal way described above. Finally, for each grouping and thus for each
chosen angle $\phi$, we compute the information $\Delta I(\phi)$ and the
entropy $\Delta H(\phi)$. In addition, we compute the angles between all pairs
of vectors in the list and plot them as a histogram approximating the
distribution function $g(\phi)$.

\subsection{A typical numerical result}

In this section, we present a typical numerical result for the quantum kicked
top taken from \cite{Schack1996b}, where more details can be found.  We look at
the time evolution of an initial Hilbert-space vector $|\psi_0\rangle$ at
discrete times $nT$. After $n$ time steps, the unperturbed vector is given by
\begin{eqnarray}
|\psi_n\rangle =\hat T^n |\psi_0\rangle\;,
\end{eqnarray}
where $\hat T$ is the unitary Floquet operator \cite{Frahm1985,Haake1987}
\begin{eqnarray}
\hat T=e^{-i(k/2J){\hat J}_x^2}e^{-i\pi{\hat J}_z/2}\;, 
\label{eqqtop}
\end{eqnarray}
and where $\hbar\hat{\bf J}=\hbar({\hat J}_x,{\hat J}_y,{\hat J}_z)$ is 
the angular momentum vector for a spin-$J$ particle evolving in
$(2J+1)$-dimensional Hilbert space. 

Depending on the initial condition, the classical map corresponding to the
Floquet operator~(\ref{eqqtop}) displays regular as well as chaotic behavior
{\cite{Haake1987}}. Following {\cite{Peres1991b}}, we choose initial
Hilbert-space vectors for the quantum evolution that correspond to classical
initial conditions located in regular and chaotic regions of the classical
dynamics, respectively. For this purpose, we use {\it coherent states\/}
\cite{Radcliffe1971,Atkins1971,Perelomov1986}. In this section, we consider two
initial states.  The first one is a coherent state centered in a regular region
of the classical dynamics; we refer to it as the {\it regular initial
state}. The second one, referred to as the {\it chaotic initial state}, is a
coherent state centered in a chaotic region of the classical dynamics.

The perturbation is modeled as an additional rotation by a small random angle
about the $z$ axis.  The system state after $n$ perturbed steps is thus given
by
\begin{equation}
|\psi_n\rangle = 
  \hat T(l_n) \cdots \hat T(l_1)\, |\psi_0\rangle \;,
\end{equation}
where $\hat T(l_m) = e^{-i g l_m \hat J_z} \hat T$, with $l_m=\pm1$, is the
unperturbed Floquet operator~(\ref{eqqtop}) followed by an additional rotation
about the $z$ axis by an angle $l_m g=\pm g$, the parameter $g$ being the 
{\it perturbation strength}. There are $2^n$ different perturbation
histories obtained by applying every possible sequence of perturbed unitary
evolution operators $\hat T(-1)$ and $\hat T(+1)$ for $n$ steps. We have
applied the method described in Sec.~\ref{secdist} to find numerically a nearly
optimal grouping of the list $\cal L$ of $2^n$ vectors generated by all
perturbation histories.

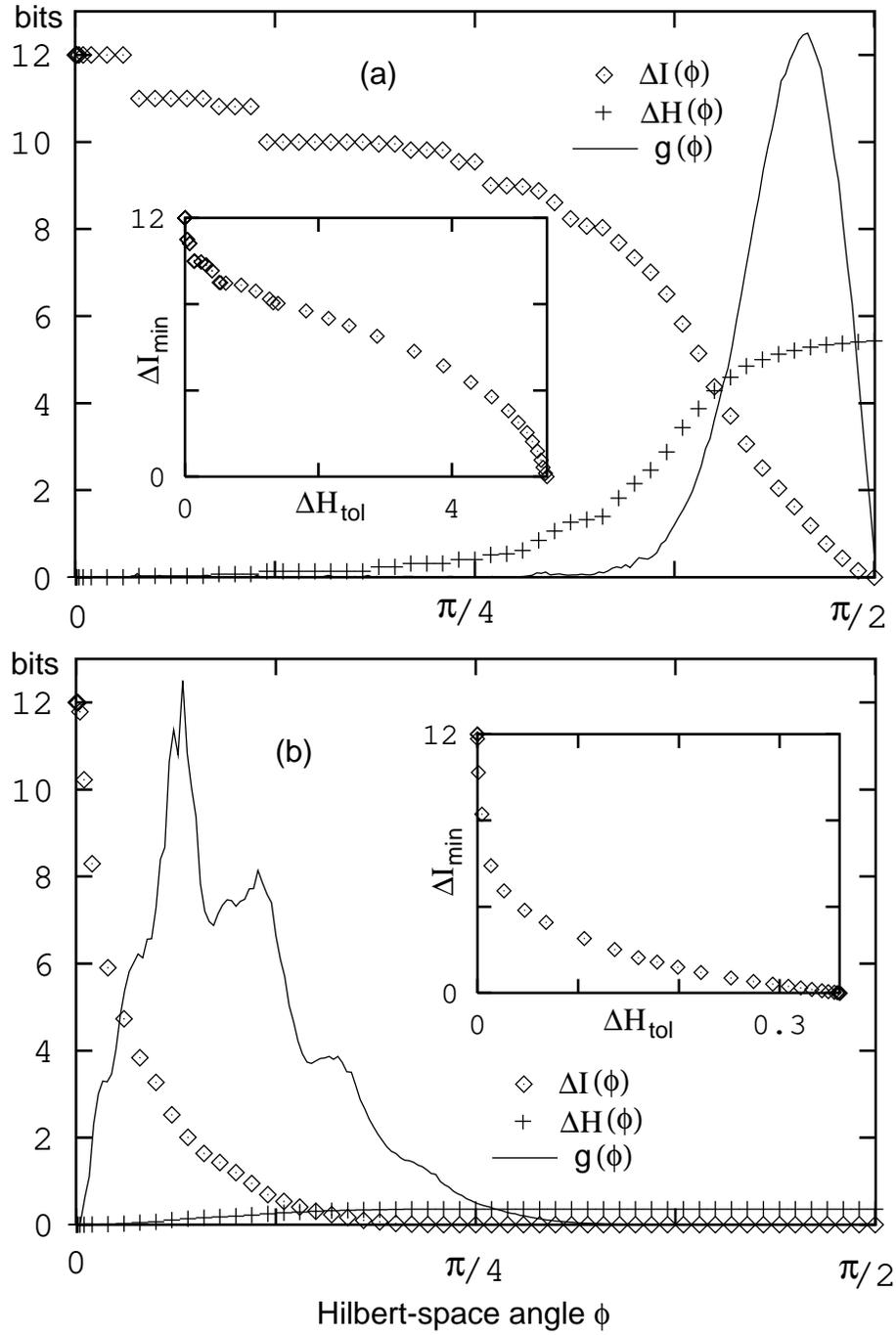
\begin{figure}
\caption{Results characterizing the distribution of Hilbert-space vectors for
the perturbed kicked top. (a) Chaotic case, (b) regular case. For details see
the text.}
\label{figtop}
\end{figure}

Figure \ref{figtop} shows results for spin $J=511.5$ and a total number of
$2^{12}=4\,096$ vectors after $n=12$ perturbed steps \cite{Schack1996b}. We
used a {\it twist parameter\/} $k=3$ and perturbation strength $g=0.003$. For
Fig.~\ref{figtop}(a), the chaotic initial state was used. The distribution of
Hilbert-space angles, $g(\phi)$, is concentrated at large angles; i.e., most
pairs of vectors are far apart from each other.  The information $\Delta I$
needed to track a perturbed vector at resolution level $\phi$ is 12 bits at
small angles, where each group contains only one vector. At $\phi\simeq\pi/16$
the information suddenly drops to 11 bits, which is the information needed to
specify one pair of vectors out of $2^{11}$ pairs, the two vectors in each pair
being generated by perturbation sequences that differ only at the first step.
The sudden drop of the information to 10 bits at $\phi\simeq\pi/8$ similarly
indicates the existence of $2^{10}$ quartets of vectors, generated by
perturbation sequences differing only in the first two steps.
Figure~\ref{figtop}(a) suggests that, apart from the organization into pairs
and quartets, there is not much structure in the distribution of vectors for a
chaotic initial state.  The $2^{10}$ quartets seem to be rather uniformly
distributed in a $n_d=46$-dimensional Hilbert space (see~\cite{Schack1996b}
for a definition of the number of explored Hilbert-space dimensions, $n_d$).

The inset in Fig.~\ref{figtop}(a) shows the approximate functional dependence
of the information needed about the perturbation, $\Delta I_{\rm min}$, on the
tolerable entropy $\Delta H_{\rm tol}$, based on the data points $\Delta
I(\phi)$ and $\Delta H(\phi)$. There is an initial sharp drop of the 
information, reflecting the grouping of the vectors into pairs and quartets. 
Then there is a roughly linear decrease of the information over a wide range 
of $\Delta H_{\rm tol}$ values, followed by a final drop with increasing slope 
down to zero at the maximum value of the tolerable entropy, 
$\Delta H_{\rm tol}=\Delta H_{\cal S}$.  The large slope of the curve near 
$\Delta H_{\rm tol}=\Delta H_{\cal S}$ can be regarded as a signature of 
hypersensitivity to perturbation. The linear regime at intermediate values 
of $\Delta H_{\rm tol}$ is due to the finite size of the sample of vectors: 
in this regime the entropy $\Delta H_r$ of the $r$th group is limited by 
$\log_2N_r$, the logarithm of the number of vectors in the group.

Figure~\ref{figtop}(b) shows data for $2^{12}$ vectors after 12 perturbed steps
in the regular case.  The distribution of perturbed vectors starting from the
regular initial state is completely different from the chaotic initial
condition of Fig.~\ref{figtop}(a).  The angle distribution $g(\phi)$ is
conspicuously nonrandom: it is concentrated at angles smaller than roughly
$\pi/4$, and there is a regular structure of peaks and valleys.  Accordingly,
the information drops rapidly with the angle $\phi$. The number of explored
dimensions is $n_d=2$, which agrees with results of Peres {\cite{Peres1991b}}
that show that the quantum evolution in a regular region of the kicked top is
essentially confined to a 2-dimensional subspace.  The $\Delta I_{\rm min}$
vs.~$\Delta H_{\rm tol}$ curve in the inset bears little resemblance to the
chaotic case. Summarizing, one can say that, in the regular case, the vectors
do not get far apart in Hilbert space, explore only few dimensions, and do not
explore them randomly.

To obtain better numerical evidence for hypersensitivity in the chaotic case
and for the absence of it in the regular case would require much larger samples
of vectors, a possibility that is ruled out by restrictions on computer memory
and time. The hypothesis most strongly supported by our data is the random
character of the distribution of vectors in the chaotic case. In the following
section we show that randomness in the distribution of perturbed vectors
implies hypersensitivity to perturbation.

\subsection{Discussion}

Guided by our numerical results we now present an analysis of hypersensitivity
to perturbation for quantum systems based on the conjecture that, for chaotic
systems, Hilbert space is explored randomly by the perturbed vectors. We
consider a Hamiltonian quantum system whose classical phase-space dynamics is
chaotic and assume the system is perturbed by a stochastic Hamiltonian that
classically gives rise to diffusion on phase space.  We suppose that at time
$t=0$ the system's state vector has a Wigner distribution that is localized on
phase space. We further assume that at $t=0$ the perturbation is just becoming
effective in the classical sense described in Sec.~\ref{secclassical}.

Our numerical analyses {\cite{Schack1993e,Schack1994b,Schack1996b}} suggest the
following picture. For times $t>0$, the entropy $\Delta H_{\cal S}$ of the
average density operator $\hat\rho_{\cal S}$ (\ref{eqrhos}) increases linearly
with time. This is in accordance with an essentially classical argument given
by Zurek and Paz \cite{Zurek1994a}. Denoting the proportionality constant by
$\kappa$, we have
\begin{equation}
\Delta H_{\cal S} \simeq \kappa t \;.
\end{equation}
Since the von Neumann entropy of a density operator is bounded by the logarithm
of the dimension of Hilbert space, it follows that the realizations of the
perturbation---i.e., the state vectors that result from the different
perturbation histories---explore at least a number
\begin{equation}
{\cal D}(t) \equiv 2^{\Delta H_{\cal S}} \simeq 2^{\kappa t} 
\label{eqd}
\end{equation}
of Hilbert-space dimensions, which increases exponentially.  Our main
conjecture now is that these dimensions are explored quasi-randomly, i.e., that
the realizations of the perturbation at time $t$ are distributed essentially
like random vectors in a ${\cal D}(t)$-dimensional Hilbert space.

Starting from this main conjecture, we will now derive an estimate of the
information $\Delta I_{\rm min}$ needed to keep the system-entropy increase
below the tolerable amount $\Delta H_{\rm tol}$. Following the discussion on
grouping vectors in Sec.~\ref{secdist}, a tolerable entropy increase $\Delta
H_{\rm tol}$ corresponds to gathering the realizations of the perturbation into
Hilbert-space spheres of radius $\phi$. The state vectors in each such sphere
fill it randomly (since the perturbation is diffusive, there are plenty of
vectors), so the entropy of their density operator---which is the tolerable
entropy---is
\begin{equation}
\Delta H_{\rm tol} = -\left(1-\frac{{\cal D}-1}{{\cal D}}\sin^2\phi\right)
\log_2\!\left(1-\frac{{\cal D}-1}{{\cal D}}\sin^2\phi\right)
\;-\frac{{\cal D}-1}{{\cal D}}\sin^2\phi\,
\log_2\!\left(\frac{\sin^2\phi}{{\cal D}}\right)
\label{eqhphi}
\end{equation}
(Eq.~(B6) of \cite{Schack1994b}). The number of spheres of radius $\phi$ in
${\cal D}$-dimensional Hilbert space is $(\sin\phi)^{-2({\cal D}-1)}$ 
(Eq.~(5.1) of \cite{Schack1994b}), so the information needed to specify a 
particular sphere is
\begin{equation}
\Delta I_{\rm min} \simeq \Delta\tilde I_{\rm min} \equiv
  \log_2\!\left( (\sin\phi)^{-2({\cal D}-1)} \right)
 = -({\cal D}-1) \log_2(\sin^2\phi) \;.
\label{eqiphi}
\end{equation}
The information $\Delta\tilde I_{\rm min}$ consistently underestimates the
actual value of $\Delta I_{\rm min}$, which comes from an optimal grouping of
the random vectors; the reason is that the perfect grouping into nonoverlapping
spheres of uniform size assumed by Eq.~(\ref{eqiphi}) does not exist.

Using Eq.~(\ref{eqiphi}) to eliminate $\phi$ from Eq.~(\ref{eqhphi}) gives an
expression for $\Delta H_{\rm tol}$ as a function of $\Delta\tilde I_{\rm min}$,
\begin{eqnarray}
\Delta H_{\rm tol}  
 &=& -\left(1- \frac{{\cal D}-1}{{\cal D}}2^{-\Delta\tilde I_{\rm min}/({\cal D}
 -1)}\right) \log_2\!\left(1-\frac{{\cal D}-1}{{\cal D}}
 2^{-\Delta\tilde I_{\rm min}/({\cal D}-1)}\right)  \nonumber\\
 &\mbox{}& \;\;\;\;\;\;-\frac{{\cal D}-1}{{\cal D}}
 2^{-\Delta\tilde I_{\rm min}/({\cal D}-1)}\, \log_2\!\left(\frac{
 2^{-\Delta\tilde I_{\rm min}/({\cal D}-1)}}{{\cal D}}\right) \;,
\label{eqhi}
\end{eqnarray}
 from which ${\cal D}$ could be eliminated in favor of $\Delta H_{\cal S}$ by
invoking Eq.~(\ref{eqd}). The behavior of $\Delta\tilde I_{\rm min}$ as a
function of $\Delta H_{\rm tol}$ expressed in Eq.~(\ref{eqhi}) is the universal
behavior that we conjecture for chaotic systems, except for when $\Delta H_{\rm
tol}$ is so close to $\Delta H_{\cal S}$ that $\Delta\tilde I_{\rm min}\alt1$,
as the spheres approximation used above breaks down for angles $\phi$ for which
Hilbert space can accommodate only one sphere. Since $\Delta H_{\rm tol}$
increases and $\Delta\tilde I_{\rm min}$ decreases with $\phi$, $\Delta\tilde
I_{\rm min}$ increases as $\Delta H_{\rm tol}$ decreases from its maximum value
of $\Delta H_{\cal S}$.

To gain more insight into Eq.~(\ref{eqhi}), we calculate the derivative
\begin{equation}
{d\Delta\tilde I_{\rm min} \over d\Delta H_{\rm tol}} = 
 - {{\cal D} \over \sin^2\phi \ln(1+{\cal D}\cot^2\phi)} \;,
\label{eqdi}
\end{equation}
which is the marginal tradeoff between between information and entropy. For
$\phi$ near $\pi/2$, so that $\epsilon=\pi/2-\phi\ll1$, the information becomes
$\Delta\tilde I_{\rm min}=({\cal D}-1)\epsilon^2/\ln2$, and the derivative
(\ref{eqdi}) can be written as 
\begin{equation}
{d\Delta\tilde I_{\rm min} \over d\Delta H_{\rm tol}} \simeq
 - {{\cal D} \over \ln(1+{\cal D}\epsilon^2)} 
= -\frac{{\cal D}}
   {\displaystyle{\ln\!\left(1+
    {{\cal D}\over{\cal D}-1}\Delta\tilde I_{\rm min}\ln2\right)}} \;.
\label{eqdiapprox}
\end{equation}

For $\Delta I_{\rm min}\agt1$, i.e., when Eq.~(\ref{eqhi}) is valid, the size
of the derivative (\ref{eqdiapprox}) is determined by ${\cal D}(t)=2^{\Delta
H_{\cal S}}\simeq2^{\kappa t}$, with a slowly varying logarithmic correction.
This behavior, characterized by the typical slope ${\cal D}(t)$, gives an {\it
exponential\/} hypersensitivity to perturbation, with the classical number of
correlation cells, ${\cal R}(t)$, roughly replaced by the number of explored
Hilbert-space dimensions, ${\cal D}(t)$.

It is a remarkable fact that the concept of perturbation cell or perturbation
correlation length (see Sec.~\ref{secclassical}) did not enter this 
quantum-mechanical discussion.  Indeed, our numerical results suggest that 
our main conjecture holds for a single correlation cell, i.e., for a 
perturbation that is correlated over all of the relevant portion of phase 
space. That we find this behavior indicates that we are
dealing with an intrinsically quantum-mechanical phenomenon.  What seems to be
happening is the following. For tolerable entropies $\Delta H_{\rm tol}\agt F$,
where $2F$ is the dimension of classical phase space as in
Sec.~\ref{secclassical}, we can regard a single-cell perturbation as perturbing
a classical system into a set of nonoverlapping densities. In a quantum
analysis these nonoverlapping densities can be crudely identified with
orthogonal state vectors. The single-cell quantum perturbation, in conjunction
with the chaotic quantum dynamics, seems to be able to produce arbitrary linear
superpositions of these orthogonal vectors, a freedom not available to the
classical system. The result is a much bigger set of possible realizations of
the perturbation.

\section{Conclusion}

This paper compares and contrasts hypersensitivity to perturbation in
classical and quantum dynamics.  Although hypersensitivity provides a
characterization of chaos that is common to both classical and quantum
dynamics, the mechanisms for hypersensitivity are different classically and
quantum mechanically.  The classical mechanism has to do with the information
needed to specify the phase-space distributions produced by the
perturbation---this is classical information---whereas the quantum mechanism
has to do with the information needed to specify the random state vectors
produced by the perturbation---this is quantum information because it relies on
the superposition principle of quantum mechanics.  Captured in a slogan, the
difference is this: {\it a stochastic perturbation applied to a classical 
chaotic system generates classical information, whereas a stochastic 
perturbation applied to a quantum system generates quantum information}.

\newpage

\end{document}